\documentclass[12pt,preprint]{aastex}

\usepackage{epsfig}

\shorttitle{Electron temperatures of PNe from He~{\sc i}}
\shortauthors{Zhang et al.}

\begin{document}

\title{Electron Temperatures of Planetary Nebulae Determined from
the He~{\sc i} Discontinuities}

\author{Y. Zhang$^{1}$,
H.-B. Yuan$^2$, C.-T. Hua$^3$, X.-W. Liu$^{2,4}$, J. Nakashima$^{1}$, and S. Kwok$^{1}$}

\altaffiltext{1}{Department of Physics, University of Hong Kong, Hong Kong; 
zhangy96@hku.hk}
\altaffiltext{2}{Department of Astronomy, Peking University, Beijing 100871, 
China}
\altaffiltext{3}{Observatoire Astronomique de Marseille-Provence Laboratoire 
d'Astrophysique de Marseille P\^ole de l'Etoile Site de Ch\^ateau-Gombert 38, 
rue Fr\'ed\'eric Joliot-Curie 13388 Marseille cedex 13, France}
\altaffiltext{4}{Kavli Institute for Astronomy and Astrophysics, Peking University, Beijing 100871, China}

\begin{abstract}

We have used the He~{\sc i} discontinuities at 3421\,{\AA} to
determine the electron temperatures, designated $T_{\rm e}$(He~{\sc i}), 
for a sample of five Galactic planetary nebulae (PNe).  We compared
$T_{\rm e}$(He~{\sc i}) with the electron temperatures derived from
the hydrogen Balmer jump at 3646\,{\AA}, designated $T_{\rm e}$(H~{\sc i}),
and found that $T_{\rm e}$(He~{\sc i}) are generally lower than
$T_{\rm e}$(H~{\sc i}). There are two possible interpretations,
a) the presence of substantial He$^{2+}$ zone, or b) the
presence of hydrogen-deficient cold clumps within diffuse
nebulae.
A series of photoionization models were constructed
to test the two scenarios.
We found that the observed $T_{\rm e}$(He~{\sc i})/$T_{\rm e}$(H~{\sc i})
discrepancies are beyond the predictions of chemically homogeneous
models. Our modelling shows that the presence of a small amount of 
hydrogen-deficient inclusions seems to be able to
reproduce the observed intensities of 
He~{\sc i} discontinuities. We stress the value of He~{\sc i} discontinuities 
in investigating nebular physical conditions. 
Albeit with some observational and technical limitations, He~{\sc i} 
discontinuities should be considered in future modelling work.

\end{abstract}


\section{Introduction}

The accurate determination of chemical abundances of planetary
nebulae (PNe) is of fundamental importance to the understanding
of the  nucleosynthesis  of low- and intermediate-mass stars
and the chemical evolution of galaxies. However,
one of the main problems in nebular astrophysics is that
the heavy element abundances derived from collisionally excited lines (CEL) 
are often lower than those derived from optical recombination lines (ORLs).
The typical ORL/CEL abundance discrepancy factor (ADF) is $\sim2$ for general 
PNe. The most extreme case to date is Hf 2-2, which has a ADF
of about 70 \citep{lb06}. Recent reviews on this problem have been presented by
\citet{liu06} and \citet{pei06} (see also the references therein).
There are two possible solutions: (1) the presence of
temperature and density variations in chemically homogeneous nebulae;
(2) the two-component nebular model with hydrogen-deficient inclusions 
embedded in the diffuse nebula. In order to test the two scenarios,
we need to further investigate  nebular physical
conditions, in particular for the regions that ORLs originate in.
For this purpose, new plasma diagnostic tools rather than
the classical CEL diagnostics are badly required.

 \citet{zhang04} applied the hydrogen Balmer jump
to determine the electron temperatures -- hereafter $T_{\rm e}$(H~{\sc i})
-- for a large sample of PNe. They found that 
$T_{\rm e}$(H~{\sc i}) is systematically lower than that derived
from the ratio of nebular to auroral lines of [O~{\sc iii}]
-- hereafter $T_{\rm e}$(O~{\sc iii}), consistent with previous
results by \citet{liu93}. 
The discrepancies between $T_{\rm e}$(H~{\sc i})
and $T_{\rm e}$(O~{\sc iii}) were first studied by \citet{peim67,peim71},
who found that temperature variations within nebulae may lead to
higher $T_{\rm e}$(O~{\sc iii}) compared to $T_{\rm e}$(H~{\sc i}). To 
quantitatively study the problem, they
defined the mean square temperature variation, $t^2$. 
Photoionization models of chemically and spatially homogeneous nebulae 
yielded typical values of $t^2$ between 0.003--0.015 \citep[see][]{pei06}.
However, the observations
by \citet{liu93} and \citet{zhang04} indicated to considerably large
temperature variations, which are far beyond the predictions of
typical photoionization models. This can be ascribed to either
additional energy inputs for chemically homogeneous nebulae
or chemical inhomogeneities.

%To further investigate nebular physical conditions,
\citet{zhang05a,zhang05b} developed a method to use
the He~{\sc i} recombination line ratios to diagnose electron temperatures
-- hereafter $T_{\rm e}$(He~{\sc i}).
%For that, high signal-to-noise spectra are required
%since recombination lines have only weak dependence on electron
%temperature. 
The comparison between $T_{\rm e}$(He~{\sc i}) and
$T_{\rm e}$(H~{\sc i}) is essential to discriminate between
the two-abundance model and chemically homogeneous nebulae
with temperature and density variations as responsible for
the CEL/ORL abundance problem since the two scenarios
predict different relations between $T_{\rm e}$(He~{\sc i}) and
$T_{\rm e}$(H~{\sc i}).
From a study of a sample of 48 PNe, \citet{zhang05a} 
(hereafter Paper~{\sc i}) found 
that $T_{\rm e}$(He~{\sc i}) is significantly lower than
$T_{\rm e}$(H~{\sc i}), in favor of the two-abundance model.
However, there are two arguable problems. \citet{pei06}
found that for some PNe the $T_{\rm e}$(He~{\sc i}) values derived from
the He~{\sc i} $\lambda\lambda$3889, 4471, and 7069 are different
with those derived from the He~{\sc i} $\lambda\lambda$6678,7281.
This might be partially caused by the uncertainties of 
the density of He$^+$ zones and/or the
optical depth effects on the He~{\sc i} $\lambda$3889.
\citet{pei06} proposed that at least 10 different He~{\sc i} lines
are required to accurately deduce these parameters.
Moreover, for high-excitation PNe the He$^+$ and H$^+$ zones are
not identical, and the presence of a substantial He$^{2+}$ zone
may partially contribute to the difference between $T_{\rm e}$(He~{\sc i})
and $T_{\rm e}$(H~{\sc i}).

These methods have also been used to determine the electron temperatures
of H~{\sc ii} regions \citep{garcia05,guseva06,guseva07}. These authors found that the
temperature differences in H~{\sc ii} regions  are not so
obvious as those found in PNe.

In this paper, we use an alternative method, fitting the
He~{\sc i} discontinuities, to determine
$T_{\rm e}$(He~{\sc i}) for five Galactic PNe. 
This method is made possible by the
availability of high signal-to-noise spectra of these nebulae resulting
in a clear detection of the helium jump. 
The current study allows us to
further investigate the relation between
$T_{\rm e}$(He~{\sc i}) and  $T_{\rm e}$(H~{\sc i}).
We examine the two-abundance models and 
photoionization models of chemically
homogeneous nebulae as the explanation of the observed
$T_{\rm e}$(H~{\sc i})/ $T_{\rm e}$(He~{\sc i}) discrepancies.
%To place tighter constraints on the models,
%He~{\sc i} recombination lines are also considered in our analysis.
This paper is a complement to the authors' work on  
recombination spectra 
as probes of physical conditions of PNe
\citep{liu93,zhang04,zhang05a,zhang05b}.

\section{Observations and data reduction}

The sample includes two low-excitation PNe and three high-excitation PNe
(see Table~\ref{tab1}). The long-slit spectra were obtained using 
the f-18 Nasmyth focus of the 2.3\,m Advanced Technology Telescope
at Siding Spring Observatory on March 2000. 
Both arms of the Double Beam Spectrograph (DBS) were used with the
1200B grating in the second order to give 
 spectral resolution of 0.3\,{\AA}/pixel.
The dichroic may be removed or replaced with a flat
mirror allowing the use of either arm independently. 
The detector was a SITE $1752\times532$ chip, giving a
 spectral coverage from 3240--8520\,{\AA}.
In order to avoid loss of light due to
atmospheric refraction each PN was observed with the spectrograph slit
aligned to the parallactic angle. Furthermore, all the observations
were carried out under the best seeing conditions, along with airmass
around 1.2 to ensure the total UV light into the spectrograph slit. 
 For each PN, the exposure time was 1500\,s.

The {\sc long92} package in 
{\sc midas}\footnote{{\sc midas} is developed and distributed by the European Southern Observatory.} 
was used for the
data reduction following the standard steps, including
bias subtraction, flat-field correction, cosmic-ray rejection, and
wavelength calibration with a helium-argon calibration lamp.
Flux calibration was conducted using the spectroscopic standards
EG\,131 and LTT\,4364.
Finally, the spectra were dereddened using 
the standard Galactic extinction law for 
a  total-to-selective extinction ratio of $R=3.1$ \citep{howarth83}
and the extinction coefficients given by \citet{cahn92}.

\section{Analysis and results}

\begin{figure*}
\begin{center}
\epsfig{file=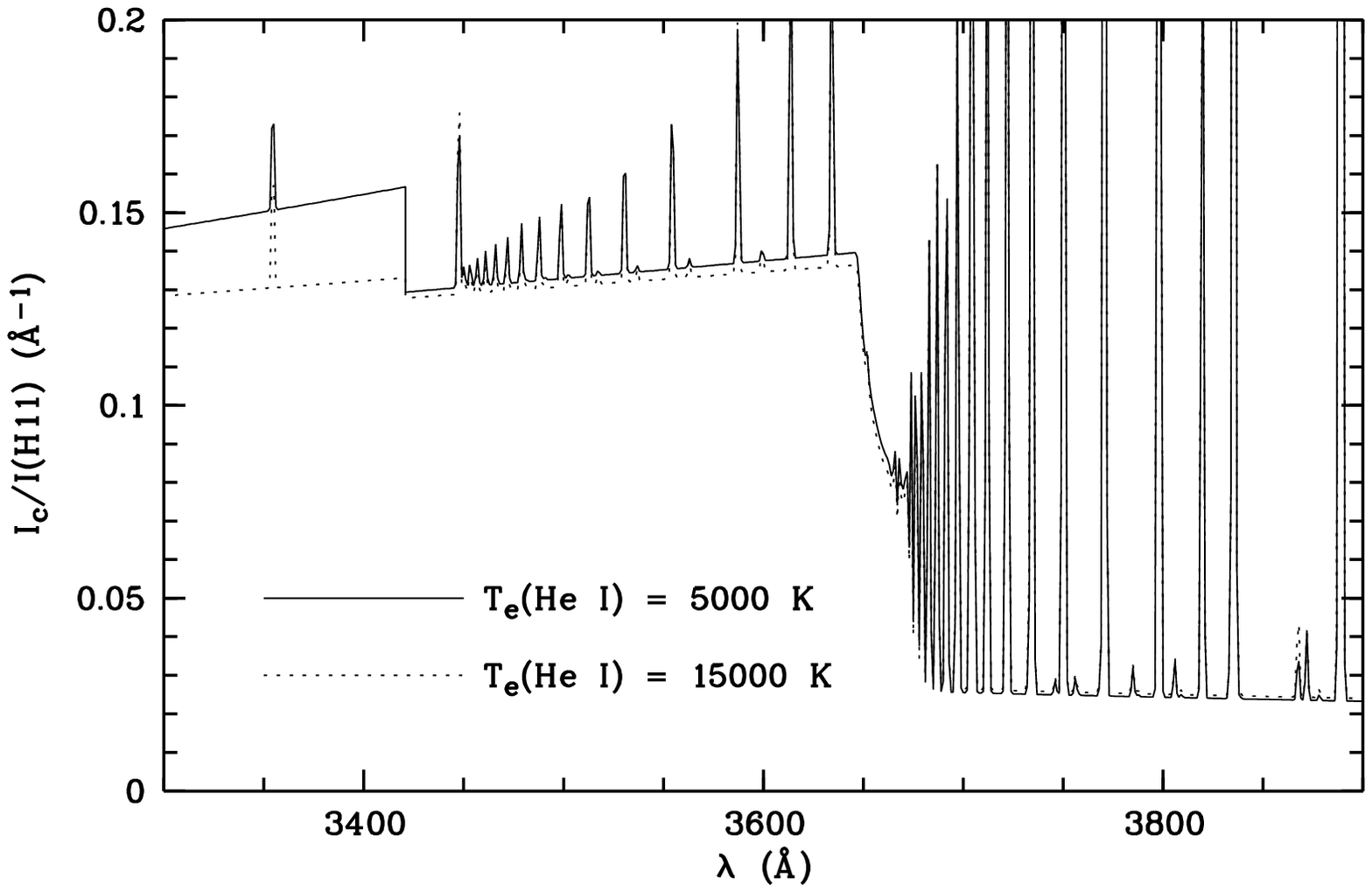,
height=10cm, bbllx=70, bblly=135, bburx=530, bbury=433, clip=, angle=0}
\end{center}
\caption{Theoretical H~{\sc i} and He~{\sc i} recombination spectra at
different $T_{\rm e}$(He~{\sc i}). For the calculations, we have
assumed $T_{\rm e}$(H~{\sc i})=10\,000\,K, He$^+$/H$^+$=0.1, and
He$^{2+}$/He$^+$=0.}
\label{theory}
\end{figure*}

The spectra allow us to simultaneously determinate $T_{\rm e}$(He~{\sc i})
and $T_{\rm e}$(H~{\sc i}) in terms of  the He~{\sc i}
discontinuities at 3421\,{\AA}
($J_{{\rm He}}=I_{3421^{-}}-I_{3421^{+}}$), produced by He$^{+}$ 
recombination to the
He~{\sc i} 2p\,$^3$P$^{\rm o}$ level, and the H~{\sc i} Balmer 
discontinuities at 3646\,{\AA}
($J_{{\rm H}}=I_{3646^{-}}-I_{3646^{+}}$). 
Note that we
do not distinguish between the electron temperatures derived
from the He~{\sc i} discontinuities and from the He~{\sc i} lines
and use $T_{\rm e}$(He~{\sc i}) to denote both of them.
The method is similar to that used to
obtain $T_{\rm e}$(H~{\sc i}) in \citet{zhang04}. To synthesize the
spectra, we consider the contributions from the free-free, free-bound,
and bound-bound emission from H~{\sc i}, He~{\sc i}, and He~{\sc ii} and the 
two-photon decay from the 2$^2S$ level of hydrogen. 
The coefficients of the H~{\sc i}, He~{\sc i}, and He~{\sc ii} continuous 
emission as functions of temperature are available in \citet{ercolano06}.
The scattering from the central star is simulated using a
modified black-body function, $I_\lambda\propto\lambda^{-\beta}B(\lambda,T)$,
where $\beta$ is the spectral index ($\beta=0$ for stardand black-body 
spectrum), $B(\lambda,T)$ is the Planck function, 
and $T$ is the temperature of the central star. 
We refer the reader to  \citet{zhang04} for a detailed description 
of our fitting procedure.  In order to illustrate our method,
Fig.~\ref{theory} gives the theoretical  spectra at different
$T_{\rm e}$(He~{\sc i}). It is clear that increasing  of
$T_{\rm e}$(He~{\sc i}) will  decrease the He~{\sc i}
discontinuities. This is a consequence of the fact that the number
of the lowest-energy free electrons, which is proportional to
the intensity of the He~{\sc i} free-bound emission at 3421$^{-}$\,{\AA},
decreases with increasing electron temperature. 
For the same reason,  the H~{\sc i} Balmer discontinuities
decrease with increasing  $T_{\rm e}$(H~{\sc i}).
Note that the $J_{{\rm H}}$ values
may slightly change with $T_{\rm e}$(He~{\sc i}) because of
a small contribution
from the He~{\sc i} discontinuity at 3680\,{\AA} produced by 
recombination to the 2\,$^1$P level.

Consequently,
$T_{\rm e}$(He~{\sc i}) and $T_{\rm e}$(H~{\sc i}) can be derived by fitting 
the theoretical to the observed spectra. Before the fitting, we
require to measure the He$^+$/H$^+$ and He$^{2+}$/He$^+$ abundance ratios.
The He$^+$/H$^+$ ratios are obtained from a few strong He~{\sc i}
lines around the He~{\sc i} discontinuities. The deduced
He$^+$/H$^+$ ratios are relatively insensitive to the adopted electron
temperatures and densities. The flux uncertainties of 
the He~{\sc i}  lines are less than 15$\%$. 
% (also note that the temperatures 
%of He~{\sc i} line emission regions might be different with 
%those of He~{\sc i} discontinuity emission regions).
The He$^{2+}$/He$^+$ ratios are taken from  \citet{cahn92} (see Table~\ref{tab1}).
The errors in temperature determination
caused by uncertainties of the He$^{2+}$/He$^+$ ratios
are negligible since the contribution from He~{\sc ii} 
to the continuum emission is much smaller than that from
H~{\sc i} and He~{\sc i}.

\begin{figure*}
\begin{center}
\epsfig{file=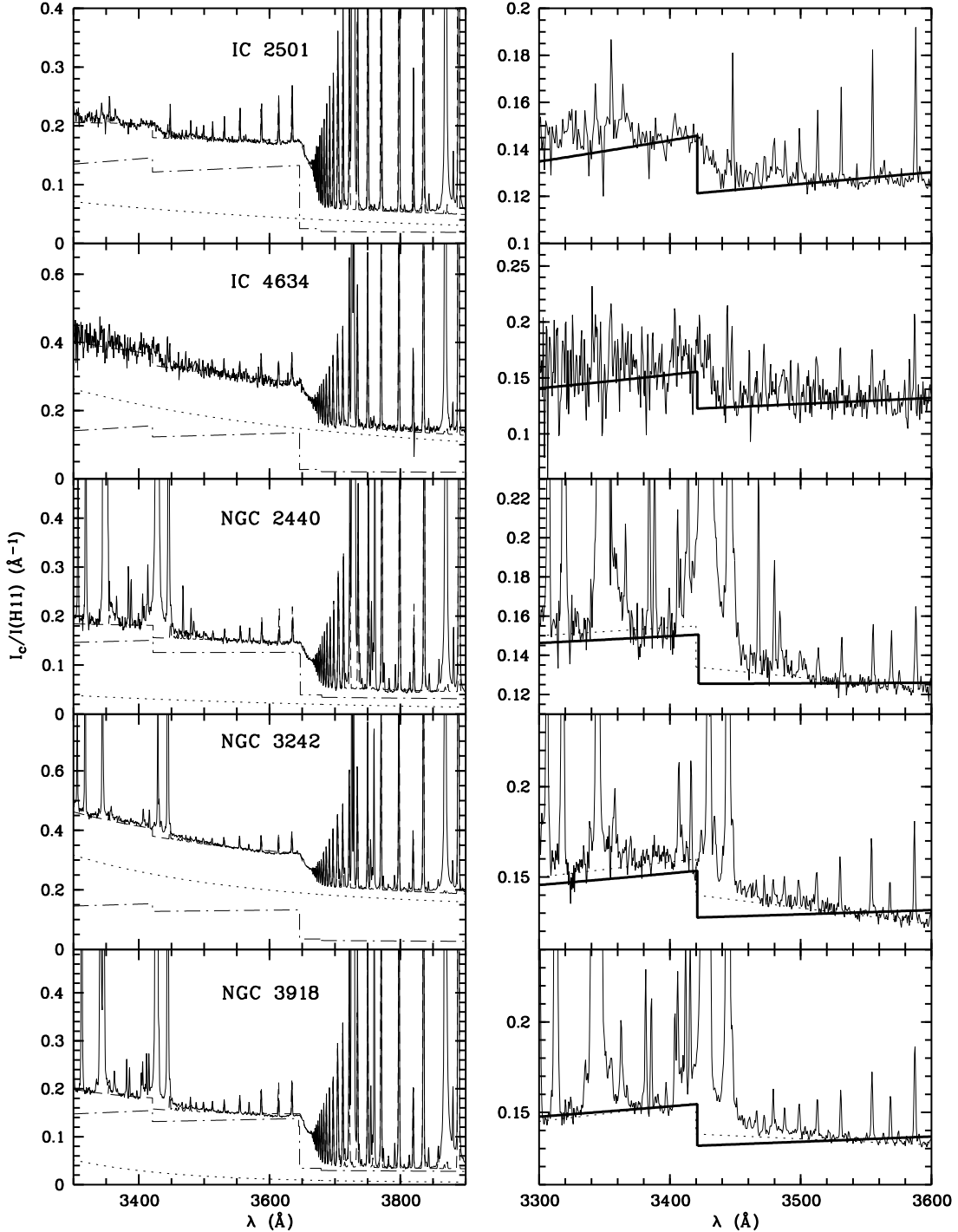,
height=19cm, bbllx=37, bblly=45, bburx=557, bbury=716, clip=, angle=0}
\end{center}
\caption{{\it Left panels:} a comparison of observed (solid lines) and theoretical (dashed
lines) spectra. The dotted-dashed lines and the dotted lines represent
the theoretical nebular continuum and scattered stellar light, respectively.
{\it Right panels:} zoom-in views of the spectra around the He~{\sc i} discontinuities at 3421\,{\AA}.
The contributions from scattered-light continuum have been subtracted.
The modelling nebular continuities are overlaid with thick solid lines. 
Note that although our models give
the best fitting to the global spectra from 3240--4000\,{\AA},
they may not satisfactorily reproduce the narrow spectral ranges of the
three high-excitation PNe from 3421--3500\,{\AA} for a variety of reasons 
(see the context). The dotted
lines denote the best fitting to the narrow spectral ranges.
}
\label{spe}
\end{figure*}

Figure.~\ref{spe} shows the fitting results for the five PNe.
In order to reduce the errors caused by uncertainties
in the reddening correction and flux calibration, we
have normalized all the spectra to H~11 $\lambda3770$.
The resultant $T_{\rm e}$(He~{\sc i}) and $T_{\rm e}$(H~{\sc i}) 
are given in Table~\ref{tab1}. Given  the relations
%$T_{\rm e}$(He~{\sc i}, H~{\sc i})$\propto J_{\rm He, H}^{-\alpha}$,
$T_{\rm e}({\rm He~{\sc I}, H~{\sc I}})\propto J_{\rm He, H}^{-\alpha}$,
where $\alpha\approx1.5$,
the errors introduced by measurement uncertainties
can be estimated through
\begin{equation}
\Delta T_{\rm e}({\rm He~I, H~I})=\frac{3\sigma}{J_{\rm He, H}}
T_{\rm e}({\rm He~I, H~I}),
\end{equation}
where $\sigma$ is the rms noise derived from the spectral
regions free of emission lines, which cover a 
wide wavelength range. We found that for all the 
nebulae the 3$\sigma$ values are well below the
$J_{\rm He, H}$ values. Another error source arises in
the uncertainties of the He~{\sc i} line emission coefficients, which
might introduce an error in the He~{\sc i} temperature determination of about
$20\%$ according to a comparison of the atomic data given
by different authors \citep{brocklehurst72,benjamin99,bauman05}.
For high-excitation PNe,
additional errors are caused by the strong O~{\sc iii} Bowen
lines around the He~{\sc i} discontinuities. In such cases,
our fittings were based on the spectral regions with wavelengths
departing from 3421\,{\AA}, introducing a $\sim 15\%$ 
errors in derived $T_{\rm e}$(He~{\sc i})
caused by uncertainties in flux calibration.
The emission efficient of  hydrogen two-photon continuity  was
deduced under Case A assumption and might be
significantly underestimated under Case B conditions due to
the collisional excitation from 2$^2P$ to 2$^2S$.
However, the excess emission of two-photon process can be 
mostly compensated by the simulated scattering continuities, and
we do not require accurate estimate of the absolution
magnitude of two-photo emission.
The uncertainties in the smooth continuities hardly affect
our results mainly based on the 3421\,{\AA} discontinuity.

The right panels of Figure.~\ref{spe} give zoom-in views of
the spectra around the He~{\sc i} discontinuities. For some PNe, 
the narrow spectral ranges from 3421--3500\,{\AA} cannot be
well fitted by our models due to the facts that
the models did not treat the extended wings of O~{\sc iii} Bowen lines and
the blended high-$J$ He~{\sc i} lines at the red side
of the 3421\,{\AA} discontinuity,
and the low sensitivity near the CCD edges leads to a larger slope. 
Nevertheless, these effects on our results should  be somewhat
insignificant since the deviations exist only in a relatively
limited wavelength range. Of course,  if our models merely take into 
account the spectra from 3300--3600\,{\AA}, we can achieve a better fitting in
such a narrow wavelength range by adjusting the contribution from scattering 
continuities. However, this will result in a failed
fitting of the global spectra. 
Modelling different spectral ranges to a reasonably good extent,
we have assessed the errors in $T_{\rm e}$(He~{\sc i}). In Figure.~\ref{spe},
we overlaid the best fitting to the narrow spectral ranges (3300--3600\,{\AA})
of NGC\,2440, NGC\,3242, and NGC\,3918, which
results in $T_{\rm e}$(He~{\sc i})$=6840$\,K, 7500\,K, and 8250\,K
for the three high-excitation PNe, respectively. We noted that
these values, which should be larger than the
real $T_{\rm e}$(He~{\sc i}) values, 
are still lower than $T_{\rm e}$(H~{\sc i}).

Three of the PNe in the recent sample have been studied
in Paper~{\sc i}.
For comparison, Table~\ref{tab1} gives $T_{\rm e}$(H~{\sc i})
and $T_{\rm e}$(He~{\sc i}) determined in Paper~{\sc i}. 
Considering the error bars, they are in agreement with the results 
derived in the recent work. Note that the $T_{\rm e}$(He~{\sc i})
in Paper~{\sc i} were derived from He~{\sc i} lines ratios and 
might possibly be different with those from He~{\sc i}
discontinuities due to the effects of radiative transfer and 
temperature inhomogeneities.

\begin{figure*}
\begin{center}
\epsfig{file=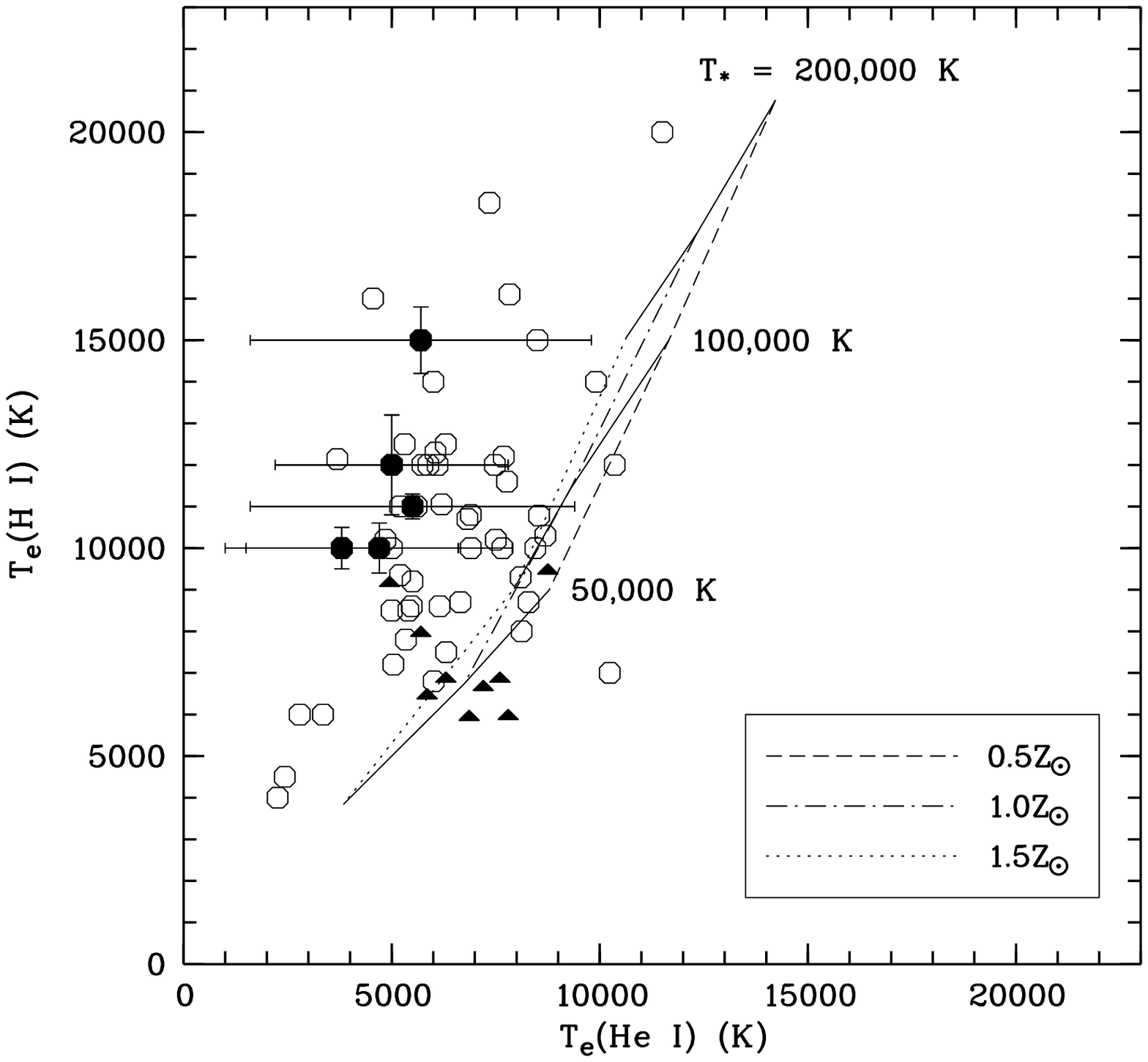,
height=12cm, bbllx=47, bblly=124, bburx=527, bbury=573, clip=, angle=0}
\end{center}
\caption{$T_{\rm e}$(H~{\sc i}) versus $T_{\rm e}$(He~{\sc i}).
The filled circles with error bars are those with $T_{\rm e}$(He~{\sc i})
determined from the He~{\sc i} discontinuities in this work. The
open circles are those with $T_{\rm e}$(He~{\sc i})
deduced from He~{\sc i} lines in Paper~{\sc i}. The filled
triangles are the results for H~{\sc ii} regions obtained by
\citet{garcia05}. These lines are the predictions by
{\sc cloudy} for different effective temperatures
of the central stars and metallicities of the nebulae
(see the context).  }
\label{com}
\end{figure*}

In Fig.~\ref{com}, we compare $T_{\rm e}$(H~{\sc i}) and
$T_{\rm e}$(He~{\sc i}). These points derived in the recent
work and those derived in Paper~{\sc i} are located at approximately
the same position in this figure. An inspection of
Fig.~\ref{com} shows that $T_{\rm e}$(He~{\sc i})
is consistently lower than  $T_{\rm e}$(H~{\sc i}),
confirming our previous results in Paper~{\sc i}.
In the scenario of the two-abundance model, the
He~{\sc i} emission are significantly contaminated
by emission from  the extremely cold hydrogen-deficient inclusions 
while the H~{\sc i} recombination spectra dominantly originate
in the normal component under a typical temperature of $T_{\rm e}\sim
10^4$~K. This provides a plausible
explanation for the lower $T_{\rm e}$(He~{\sc i}) compared to
$T_{\rm e}$(H~{\sc i}). In Fig.~\ref{com},
we also overplot the results for 
H~{\sc ii} regions obtained by \citet{garcia05}, which are
located at a more compact area in this figure and indicate
to a smaller $T_{\rm e}$(He~{\sc i})/$T_{\rm e}$(H~{\sc i})
discrepancy in H~{\sc ii} regions compared to PNe.

\section{Discussion}

\subsection{He~{\sc i} lines}

A few strong He~{\sc i} lines have been clearly detected in the red arm.
They allow an independent examination of our results derived from the
He~{\sc i} discontinuities.  This is non trivial in that
density inhomogeneities and radiative transfer effects 
could affect our results to some extent.

\begin{figure*}
\begin{center}
\epsfig{file=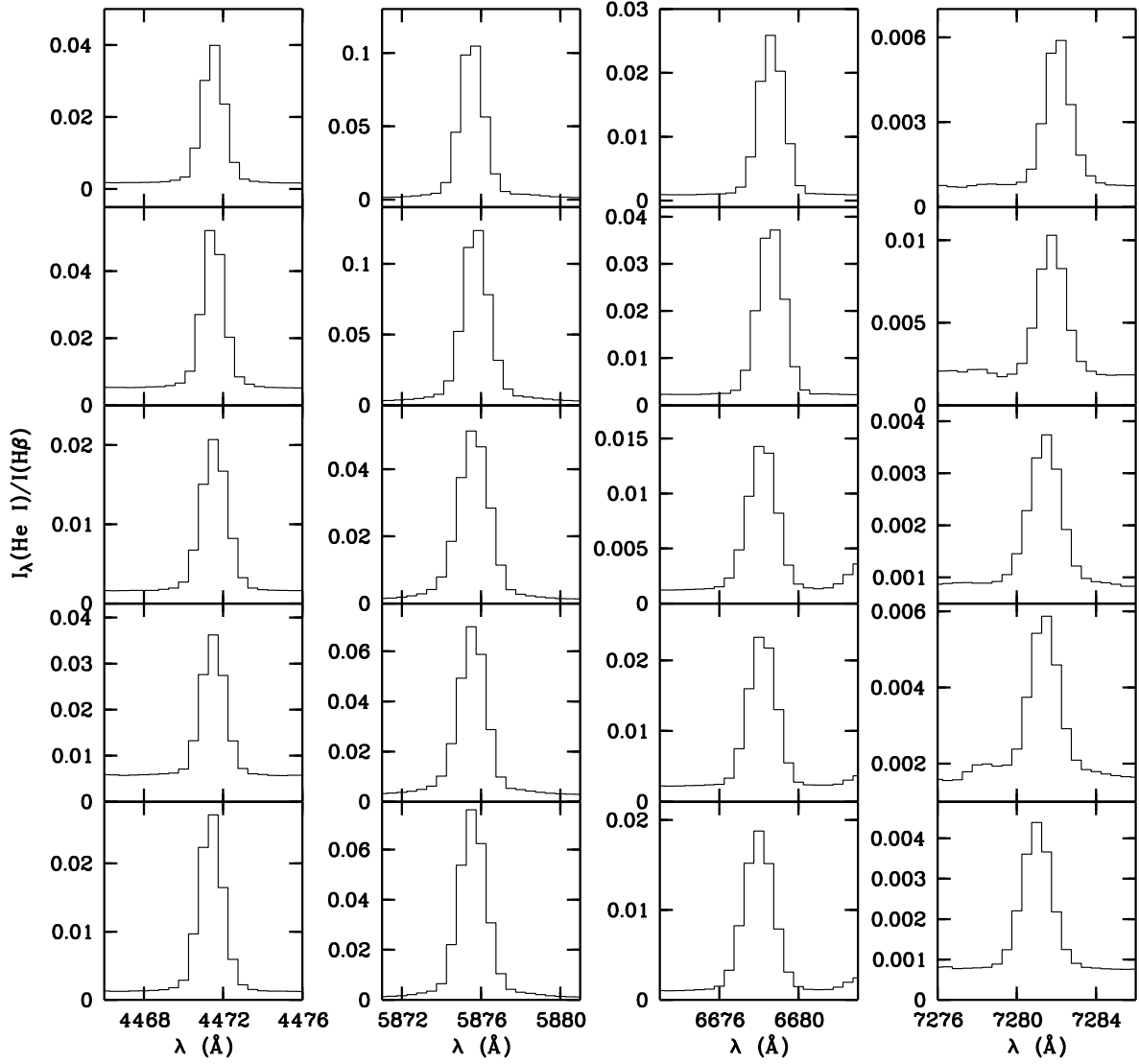,
height=15cm, bbllx=36, bblly=41, bburx=527, bbury=505, clip=, angle=0}
\end{center}
\caption{He~{\sc i} 4471, 5876, 6678, 7281 lines (from left to right) detected
in IC~2501, IC~4634, NGC~2240, NGC~3242, and NGC~3918 (from top to bottom). 
}
\label{obsline}
\end{figure*}

\begin{figure*}
\begin{center}
\epsfig{file=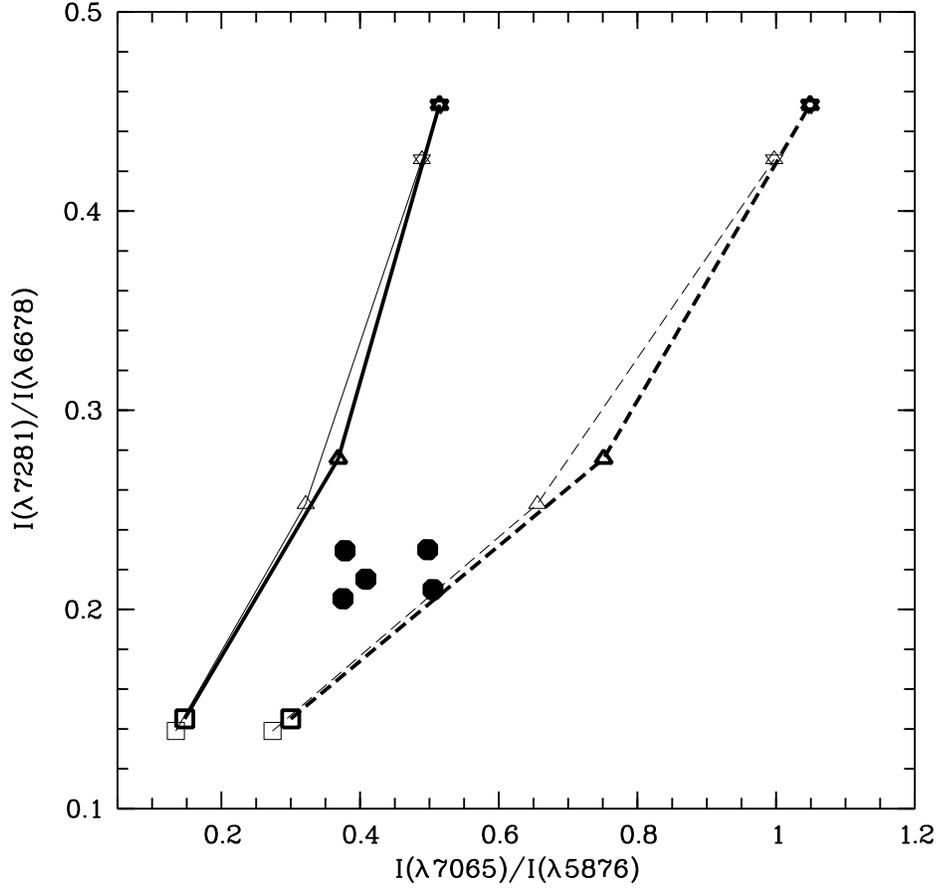,
height=12cm, bbllx=47, bblly=124, bburx=536, bbury=573, clip=, angle=0}
\end{center}
\caption{ Theoretical predictions of the He~{\sc i} line ratios
$I(\lambda7281)/I(\lambda6678)$ versus $I(\lambda7065)/I(\lambda5876)$,
derived at different physical conditions: $T_{\rm e}=5000$\,K (squares),
10\,000\,K (triangles), and 20\,000,\,K (stars); $N_{\rm e}=10^{^4}$\,cm$^{-3}$ (light lines) and
$10^{^6}$\,cm$^{-3}$ (thick lines); $\tau_{3889}=0$ (solid lines) and 10 (dashed lines).
The filled circles represent the observations.
 }
\label{heline}
\end{figure*}

Here we took into account the four strongest He~{\sc i} lines in 
the red arm, $\lambda\lambda7281$, 6678, 7065, and 5876, as
shown in Fig.~\ref{obsline}. Their intensities are listed
in Table~\ref{tab2}. In Paper~{\sc i}, we have shown that
the He~{\sc i} ratio $I(\lambda7281)/I(\lambda6678)$ provides
an useful temperature diagnostic. According to
\citet{benjamin02}, the He~{\sc i} $\lambda\lambda7065,5876$ lines
are from triplet states and are essentially affected by the optical
depth of the 2s$^3$S level.  Fig.~\ref{heline} gives the theoretical
predictions of the He~{\sc i} ratio $I(\lambda7281)/I(\lambda6678)$
versus He~{\sc i} ratio $I(\lambda7065)/I(\lambda5876)$ at
different electron temperatures, electron densities ($N_{\rm e}$), 
and optical depths of the He~{\sc i}
2s$^3$S--3p$^3$P$^{\rm o}$$\lambda$3889 line ($\tau_{3889}$).
The observed values for our sample are also shown in this figure.

The diagram is useful to investigate nebular physical conditions.
An inspection of Fig.~\ref{heline} shows that the
$T_{\rm e}$(He~{\sc i}) values
are consistently lower than 10\,000\,K, which represents
the lowest value of $T_{\rm e}$(H~{\sc i}) in our sample, 
supporting the conclusion drawn from the He~{\sc i} discontinuities. 
We also found that the $\tau_{3889}$ values range from 0 to 10.
These line ratios have only weak dependence on electron densities and
are hardly used to determined $N_{\rm e}$(He~{\sc i}).

Table~\ref{tab2} gives the $T_{\rm e}$(He~{\sc i}) values
derived from the He~{\sc i} line observations. The uncertainties
of electron densities and optical opacities introduce an error
of about 1500\,K. Using the spectral lines
falling in the overlapping wavelength region of two adjacent orders,
we estimated the uncertainties in flux calibration to be about
5$\%$. The combined errors of the resultant
$T_{\rm e}$(He~{\sc i}) amounts to about 2200\,K.
The $T_{\rm e}$(He~{\sc i}) deduced in this paper are 
biased to larger values compared to those in 
Paper~{\sc i} (although they are marginally
consistent with each other considering the error bars). This
is partly attributed to the fact that  in the present observation
the slit was put in a position closer to the center. However,
the primary cause might be that the uncertainties in 
flux calibration of the red arm are probably larger than
expect. We performed a comparison of the present 
spectra and our previous long-slit spectra  and found
that the former have systematical higher line fluxes in very red 
wavelength regions. Therefore, we suppose that
the fluxes of the He~{\sc i} $\lambda7281$ might have been
significantly overestimated, and the $T_{\rm e}$(He~{\sc i}) in
Table~\ref{tab2} should be treated as upper limits, which are
still lower than $T_{\rm e}$(H~{\sc i}). A detailed discussion of 
He~{\sc i} lines requires more precise observations in the red arm
and is beyond the focus of the present paper.

One should bear in mind that
the electron temperatures derived from the He~{\sc i} discontinuities
and those from the He~{\sc i} line ratios are not necessary identical
since they have different temperature dependences, and thus weight
different regions if large temperature fluctuations exist. The comparison
of He~{\sc i} discontinuity- and line-temperatures 
is a potentially intriguing subject for future research.

\subsection{Model tests}

An issue is invited, whether the large 
$T_{\rm e}$(H~{\sc i})/ $T_{\rm e}$(He~{\sc i}) discrepancies
in high-excitation PNe
can be ascribed to the presence of substantial He$^{2+}$ zone
which has a higher temperature than He$^{+}$ zone. In order to investigate the 
problem, we first compare the observed results with the predictions
by 1D photoionization models of chemically homogeneous 
nebulae. 
We then construct 3D photoionization models to
test the two-abundance model as the explanation of
the temperature discrepancies.

\subsubsection{Chemically homogeneous model}

To date, the published photoionization models for certain PNe
with a chemically homogeneous medium  predict only minor
discrepancy between $T_{\rm e}$(He~{\sc i}) and
$T_{\rm e}$(H~{\sc i}).  For example,
the classic photoionization model of
\object{NGC\,7662} constructed by \citet{harrington}
yields $T_{\rm e}$(H~{\sc i})$-T_{\rm e}$(He~{\sc i})$\approx 600$\,K.
To investigate the case of general PNe, we use the {\sc cloudy} code 
\citep{ferland98} to construct a series of 1D photoionization modellings 
of chemically homogeneous nebulae.
In order to compare the modelling results with the observed ones,
we define modelled $T_{\rm e}$(He~{\sc i}) and
$T_{\rm e}$(H~{\sc i}) by
\begin{eqnarray}
\nonumber
T_{\rm e}({\rm He~I, H~I})&=& \frac{\int T_{\rm e}(r) J_{\rm He, H}(r)dr}{\int 
J_{\rm He, H}(r)dr}\\
&\approx &\frac{\int T_{\rm e}(r)N_{\rm e}(r)N_{{\rm He}^+, {\rm H}^+}(r)T_{\rm e}(r)^{-2/3}dr}
{\int N_{\rm e}(r)N_{{\rm He}^+, {\rm H}^+}(r)T_{\rm e}(r)^{-2/3}dr},
\end{eqnarray}
where $r$ is the position vector, $N_{\rm e}(r)$ is the electron
density, and $N_{{\rm He}^+, {\rm H}^+}(r)$ is the
number density of He$^+$ or H$^+$ atoms.
Then $T_{\rm e}$(He~{\sc i}) and $T_{\rm e}$(H~{\sc i}) can be
simulated from the modelled
$T_{\rm e}(r)$, $N_{\rm e}(r)$, and $N_{{\rm He}^+, {\rm H}^+}(r)$.
The temperatures defined in Eq.~(2) are similar to those derived
from the observed He~{\sc i} and H~{\sc i} discontinuities.

For the modellings, a black-body 
spectral energy distribution with a luminosity
$L_*=10^{38}$\,erg\,s$^{-1}$ is assumed for the central star.
We assume that
the nebula has a homogeneous
density structure with a hydrogen density of 
$n(\rm H)=10^3$\,cm$^{-3}$. The assumed $n(\rm H)$ represents 
the lower limit of typical PNe. At a given UV radiation field,
increasing $n(\rm H)$ will decrease
the ionization parameter and the He$^{2+}$/He$^+$ ratio, and thus
decrease the contribution of He$^{2+}$ zone to the
$T_{\rm e}$(He~{\sc i})/$T_{\rm e}$(H~{\sc i}) discrepancies.

Our modellings reproduce the $T_{\rm e}$(He~{\sc i}) and 
$T_{\rm e}$(H~{\sc i}) values
in different nebular metallicities ($Z$) and 
effective temperature of the ionizing star ($T_*$). The results are
illustrated in Fig.~\ref{com}. These modellings show that
the presence of He$^{2+}$ zone can
cause a higher $T_{\rm e}$(H~{\sc i}) compared to
$T_{\rm e}$(He~{\sc i}). The $T_{\rm e}$(H~{\sc i})/$T_{\rm e}$(He~{\sc i})
discrepancies sharply increase with increasing
$T_*$ and have only weak dependency on $Z$.
Nevertheless, the large differences between
the observed $T_{\rm e}$(He~{\sc i}) and $T_{\rm e}$(H~{\sc i})
are obviously beyond the model predictions although the claim is 
probably not applied for H~{\sc ii} regions.  An inspection
of Fig.~\ref{com} suggests that chemically homogeneous models
are unable to reproduce the extremely low $T_{\rm e}$(He~{\sc i})
measured in PNe.
The observed $T_{\rm e}$(H~{\sc i}) values
can be well reproduced by using appropriate $T_*$ and $Z$.
However, in order to match the observed $T_{\rm e}$(He~{\sc i}) for 
most PNe,
we require extremely high $T_*$ ($\gg200,000$\,K) and $Z$ ($\gg1.5$\,$Z_\odot$),
which are unlikely real for our sample.

Our models do not include the  photoelectric heating by dust grains,
which can efficiently enhance the temperature in the nebular regions
close to the central star \citep{stasinska01}. In addition, there
might be other unexpected heating sources that lead to greater
temperature fluctuations. It is too early to conclude with
certainty that chemical inhomogeneities must be introduced
to explain the observed temperature discrepancies. In any case,
a more comprehensive photoionization model including extra
heating mechanisms should take into account the He~{\sc i} discontinuities.

\subsubsection{The two-abundance model}

\begin{figure*}
\begin{center}
\epsfig{file=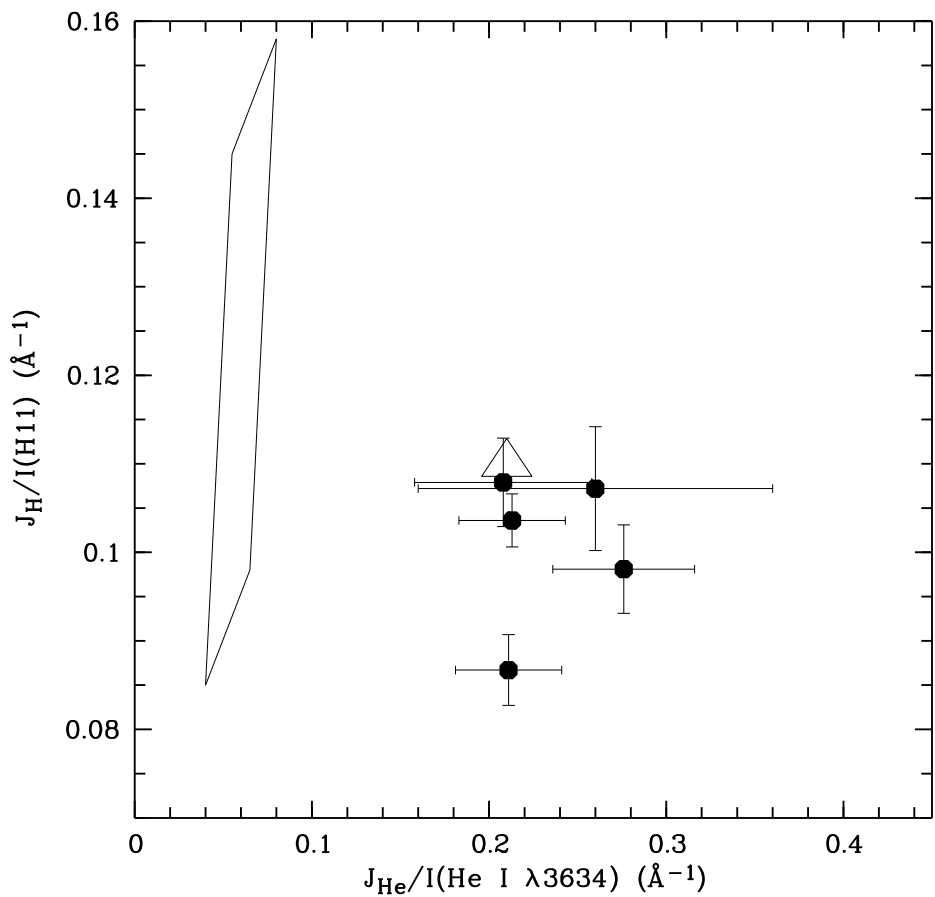,
height=12cm, bbllx=47, bblly=124, bburx=527, bbury=573, clip=, angle=0}
\end{center}
\caption{$J_{\rm H}/I$(H11) versus $J_{\rm He}/I$(He~{\sc i} $\lambda3634$).
The filled circles denote the observed values.
The open triangle is the prediction of
a two-abundance model (see the text). The enclosed region represents
the zone occupied by the predictions of chemically homogeneous models in the
parameter ranges $50,000$\,K$<T_*<200,000$\,K,
$500$\,cm$^{-3}$$<n({\rm H})<8000$\,cm$^{-3}$, and
$0.6Z_{PN}<Z<1.6Z_{PN}$.
}
\label{model}
\end{figure*}

To test the two-abundance model, we use the 3D photoionization 
code {\sc mocassin} \citep{ercolano03a}, 
which is capable of treating nebular geometric
asymmetries, density and composition inhomogeneities as well as the diffuse
radiation fields self-consistently, and thus
enables to more rigorously
study the physical properties and chemical composition of the
postulated hydrogen-deficient component compared to 1D model. 
A series of two-abundance models are constructed and the 
reproduced $J_{\rm H}$ and $J_{\rm He}$ are compared with
the observed values. We found that the presence of
a very small amount of hydrogen-deficient material is enough
to interpret the observations. An example is given in Fig.~\ref{model},
where we compare the normalized $J_{\rm H}$ and $J_{\rm He}$
observed for our PN sample and those
predicted by one of the {\sc mocassin}
models. For this two-abundance model, we have assumed
that the diffuse nebula  has a spherical structure,
 a $n$(H) value of 10$^3$\,cm$^{-3}$,
and a metallicity of $Z_{PN}$, where $Z_{PN}$ is the average metallicity
of the Galactic PNe \citep{kingsburgh94}.
The hydrogen-deficient inclusions have a filling factor
of $0.18\%$ and a $n$(H) value of 70\,cm$^{-3}$,
and are enhanced by a factor of 200 in He and by a factor
of 800 in CNONe compared to the diffuse gas. 
An  effective temperature of 120,000\,K is assumed for
the central star.
According to our modelling, 
the electron temperature of the
hydrogen-deficient inclusions is about 500\,K, which is much lower than
the average temperature of the diffuse gas of 10,770\,K. The cold
hydrogen-deficient inclusions contribute
about $45\%$ and $0.2\%$ of He~{\sc i} and H~{\sc i} line emission
intensities,
respectively. As shown in Fig.~\ref{model}, the model result
can closely match the observed $J_{\rm H}$ and $J_{\rm He}$.

For comparison, we construct a series of chemically homogeneous
models using {\sc mocassin} in the
parameter ranges $50,000$\,K$<T_*<200,000$\,K,
$500$\,cm$^{-3}$$<n({\rm H})<8000$\,cm$^{-3}$, and
$0.6Z_{PN}<Z<1.6Z_{PN}$. The predicted region
of $J_{\rm H}$ and $J_{\rm He}$ is given in Fig.~\ref{model}.
Again it is confirmed that even for high-excitation PNe
chemically homogeneous
models are unable to reproduce the observed He~{\sc i} discontinuities.

The two-abundance model can provide a natural solution for the 
observed $T_{\rm e}$(He~{\sc i})/$T_{\rm e}$(H~{\sc i}) discrepancies.
Nevertheless, we have noted some problems in our two-abundance modelling.
In order to explain the low $T_{\rm e}$(He~{\sc i}), we
require to assume a low density and high enhancement factor of heavy
elements for the hydrogen-deficient inclusions to get effective cooling.
Such a low-density and low-temperature component is unstable inside
the nebulae. Moreover, our models yield an  ADF of 10--20, higher than 
the typical value for general PNe. Probably, these imply to the presence
of some additional coolants inside hydrogen-deficient inclusions.

The hydrogen-deficient inclusions postulated
in the current study have similar properties to Abell 30's knots,
which show extremely He-rich \citep{wesson03,ercolano03b}. However,
the properties of the hydrogen-deficient inclusions in different 
PNe are probably not identical. 
A complete understanding of the
hydrogen-deficient component requires 
detailed modelling and precise measurements of heavy-element ORLs
for individual PNe, which is beyond the scope of this paper.
Our results show that the observations of He~{\sc i} 
discontinuities can provide some constraints on future study of
the hydrogen-deficient inclusions.

\section{Conclusions}

Based on a high signal-to-noise spectroscopy of a sample
of PNe, we explored
the possibility of using the He~{\sc i}
discontinuities at 3646\,{\AA} as probe of nebular physical conditions.
$T_{\rm e}$(He~{\sc i}) is found to be systematically lower than 
$T_{\rm e}$(H~{\sc i}), confirming the results of Paper~{\sc i}, in which
we determined $T_{\rm e}$(He~{\sc i}) using the He~{\sc i}
recombination lines.  
We have investigated the contribution of He$^{2+}$ zone on the
$T_{\rm e}$(He~{\sc i})/$T_{\rm e}$(H~{\sc i}) discrepancies
by constructing a series of photoionization models.
We have shown that the chemically homogeneous models predict
too small temperature discrepancies as compared to the observations.
The observations, however, can be naturally interpreted if one assumes 
the presence
of a small amount of hydrogen-deficient cold inclusions within the diffuse 
nebula.
Our results thus provide a possible evidence in favor of the two-abundance
model as the explanation of the CEL/ORL abundance problem. Nevertheless,
we cannot completely rule out the scenario of temperature fluctuations
if some extra heating mechanisms exist.

To completely understand the properties of the postulated
small hydrogen-deficient inclusions,
we need to construct detailed 3D photoionization models with very fine
mesh grids and take account of the ORLs from helium and heavy elements.
For that, an enormous amount of computer time is required.
The detailed two-abundance models for individual objects will be reported 
in forthcoming papers (e.g. Yuan et al. in preparation). 

In the current
paper, we show that the He~{\sc i} discontinuities,
notwithstanding some observational and technical limitations,
provide a valuable probe of nebular physical conditions
and can place a tight constraint for the modelling. The method
proposed in this paper is particularlly appropriate for
studying low-excitation PNe and H~{\sc ii} regions that
are free from the effects of strong O~{\sc iii} Bowen lines.
Instrument development and atomic data improvement  
will be invaluable to provide more robust conclusion.

\acknowledgments

The authors are grateful to M. A. Dopita and B. Rocca-Volmerange
for their help in ensuring successful observations. 
We thank the anonymous referee for useful comments.
CTH is grateful to HKU for having made possible the present output.
The observations were made in the frame of the
Programme International de Coop\'eration Scientifique PICS France-Australie.
The 3D modelling was carried out on the SGI Altix330 System at Department 
of Astronomy, PKU and the HPCPOWER cluster of HKU. JN acknowledges financial 
support from Seed Funding Programme for Basic Research in HKU (200802159006).
The work was partially supported by the Research Grants Council of
the Hong Kong under grants HKU7028/07P and HKU7033/08P.

\newpage

\begin{table*}
%\centering \begin{minipage}{140mm}
\caption{$T_{\rm e}$(He~{\sc i}) and $T_{\rm e}$(H~{\sc i}).
 \label{tab1}}
\begin{center}
  \begin{tabular}{lclllll}
\hline
Source & He$^{2+}$/He$^+$$^a$&\multicolumn{2}{c}{$T_{\rm e}$(He~{\sc i})$^b$ (K)}& &
\multicolumn{2}{c}{$T_{\rm e}$(H~{\sc i}) (K)}\\
\cline{3-4}\cline{6-7}
& & This work & Paper~{\sc i} & & This work & Paper~{\sc i}\\
\hline
IC 2501  & 0.002& 4700$\pm$3200 & ...        & &10000$\pm$600&  ...\\
IC 4634  & 0.000& 3800$\pm$2800 & 5000$\pm$1000 & &10000$\pm$500& 8500$\pm$400\\
NGC 2440 & 0.401& 5700$\pm$4100$^c$ & ...        & &15000$\pm$800&  ... \\
NGC 3242 & 0.230& 5000$\pm$2800$^d$ & 4500$\pm$1000 & &12000$\pm$1200& 10200$\pm$1000\\
NGC 3918 & 0.303& 5500$\pm$3900$^e$ & 6000$\pm$1000 & &11000$\pm$300 & 12300$\pm$1000\\
\hline
\end{tabular}
\begin{description}
\item $^{a}$ From \citet{cahn92}.
\item $^{b}$ Note that the $T_{\rm e}$(He~{\sc i}) values given by
Paper~{\sc i} are derived from He~{\sc i} line ratios.
\item $^{c}$ If merely the spectral range from 3300--3600\,{\AA} is taken into account,
we derived a value of 6840\,K, which represents the  upper limit of $T_{\rm e}$(He~{\sc i})
(see the dotted lines in the rigth panel of Figure~\ref{spe}).
\item $^{d}$ The upper limit is 7500\,K. Details are as indicated in Footnote $c$.
\item $^{e}$ The upper limit is 8250\,K. Details are as indicated in Footnote $c$.
\end{description}
\end{center}
\end{table*}

\begin{table*}
%\centering \begin{minipage}{140mm}
\caption{He~{\sc i} line intensities relative to $I({\rm H}\beta)=1$
and $T_{\rm e}$(He~{\sc i}) derived from He~{\sc i} lines.
 \label{tab2}}
\begin{center}
 \begin{tabular}{lccccc}
\hline
Transition   & IC 2501  & IC 4634  & NGC 2440 & NGC 3242 & NGC 3918 \\
\hline
\multicolumn{6}{l}{\bf  Singlets} \\
$\lambda$6678 & 0.0400 & 0.0590 & 0.0244 & 0.0381 & 0.0331\\
$\lambda$7281 & 0.0086 & 0.0142 & 0.0056 & 0.0082 & 0.0068\\
\hline
\multicolumn{6}{l}{\bf  Triplets} \\
$\lambda$5876 & 0.1790 & 0.2125 & 0.1039& 0.1210 & 0.1367\\
$\lambda$7065 & 0.0904 & 0.1057 & 0.0393 & 0.0494 & 0.0513\\
\hline
$T_{\rm e}$(He~{\sc i})$^a$  & 7500\,K & 8400\,K & 8000\,K & 7400\,K & 7100\,K \\
\hline
\end{tabular}
\begin{description}
\item $^{a}$ Bears an error of $\sim2200$\,K.
\end{description}
\end{center}
\end{table*}

\end{document}